\documentclass[twocolumn,amsmath,amssymb]{revtex4}

\usepackage{gensymb}
\usepackage{graphicx}% Include figure files
\usepackage{dcolumn}% Align table columns on decimal point
\usepackage{bm}% bold math
\newcommand{\etal}{\emph{et al.}}
\newcommand{\be}{\begin{equation}}
\newcommand{\ee}{\end{equation}}
\newcommand{\bfig}{\begin{figure}}
\newcommand{\efig}{\end{figure}}

\begin{document}      
\title{Anomalous Hall Effect in ZrTe$_5$} 
 
\author{Tian Liang$^{1,3}$}
\author{Jingjing Lin$^{1}$}
\author{Quinn Gibson$^{2}$}
\author{Minhao Liu$^1$}
\author{Wudi Wang$^{1}$}
\author{Hongyu Xiong$^{3}$}
\author{Jonathan A. Sobota$^{3,4}$}
\author{Makoto Hashimoto$^{5}$}
\author{Patrick S. Kirchmann$^{3}$}
\author{Zhi-Xun Shen$^{3}$}
\author{R. J. Cava$^2$}
\author{N. P. Ong$^{1}$}

\affiliation{
Departments of Physics$^1$ and Chemistry$^2$, Princeton University, Princeton, NJ 08544\\
$^3$Geballe Laboratory for Advanced Materials, Departments of Physics and Applied Physics, Stanford University, Stanford, CA 94305 and Stanford Institute for Materials and Energy Sciences, SLAC National Accelerator Laboratory, 2575 Sand Hill Road, Menlo Park, CA 94025\\
$^4$Advanced Light Source, Lawrence Berkeley National Laboratory, Berkeley, California 94720\\
$^5$Stanford Synchrotron Radiation Lightsource, SLAC National Accelerator Laboratory, 2575 Sand Hill Road, Menlo Park, California 94025
} 

%\date{\today}      
\begin{abstract}
{ZrTe$_5$ has been of recent interest as a potential Dirac/Weyl semimetal material. Here, we report the results of experiments performed via in-situ 3D double-axis rotation to extract the full $4\pi$ solid angular dependence of the transport properties. 
A clear anomalous Hall effect (AHE) was detected for every sample, with no magnetic ordering observed in the system to the experimental sensitivity of torque magnetometry. Interestingly, the AHE takes large values when the magnetic field is rotated in-plane, with the values vanishing above $\sim 60$~K where the negative longitudinal magnetoresistance (LMR) also disappears. This suggests a close relation in their origins, which we attribute to Berry curvature generated by the Weyl nodes.  
}
\end{abstract}
 
 % Produces the title
\maketitle

The notion of topology has expanded to include the Dirac/Weyl semimetals, which feature 3D Dirac states protected by symmetry.
In this context, ZrTe$_5$ has recently attracted considerable attention, following the observation of negative longitudinal magnetoresistance (LMR)~\cite{LiZrTe5}, whose origin is claimed to come from the chiral anomaly~\cite{Adler,BellJakiw,NielsenNinomiya,Xiong2015,Max2016}, reminiscent of Dirac/Weyl semimetals~\cite{Murakami2008,Murakami2014,Hosur2013,Ashvin,Ran,Kane,Liang2015,YLChen2015,HongDing2015}. Despite the observation of the negative LMR, however, unlike Na$_3$Bi~\cite{WangNa3Bi} and Cd$_3$As$_2$~\cite{WangCd3As2}, there are no theoretical predictions showing that ZrTe$_5$ is 3D Dirac/Weyl semimetal. Furthermore, the results of angle-resolved photoemission spectroscopy (ARPES) experiments~\cite{DingZrTe5,ZhangZrTe5,ManzoniZrTe5,YLChen2016,ShenZrTe5,LiZrTe5} have not converged yet.

It is therefore of interest to investigate other unusual transport properties of ZrTe$_5$, especially the anomalous Hall effect (AHE). The expression for the AHE of Dirac/Weyl semimetals is: \cite{Ran,Hosur2013,Burkov}, viz.,
\begin{eqnarray}
	\sigma_{\mathrm{AHE}} = \frac{e^2}{2\pi h}~\left|\sum \Delta \bm{k}_i\right|
\end{eqnarray}
with $\Delta \bm{k}_i$ representing the separation between the Weyl nodes in $\bm{k}$-space. In Dirac/Weyl semimetals, no ferromagnetism is required to observe the AHE because of the strong
Berry curvature $\Omega_{\bm{k}}$ produced by the Weyl nodes. To characterize ZrTe$_5$ in more detail, in this work, full 
$4\pi$ solid angular dependence of the anomalous Hall signals, i.e. the Berry curvature $\Omega_{\bm{k}}$, obtained by
in-situ 3D double axis rotation measurements, is reported.

\begin{figure*}[t]
\includegraphics[width=18 cm]{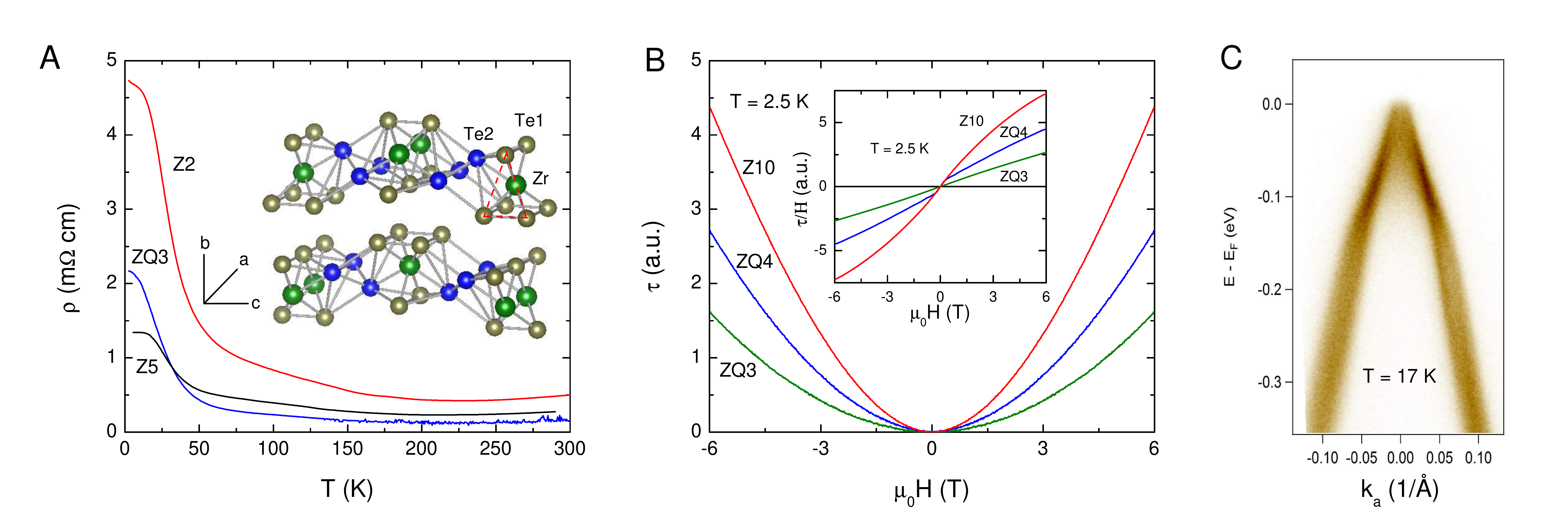}
\caption{\label{RT} 
Panel A: Resistivity as a function of temperature for selected samples. The resistivity decreases as the system cools down at high temperatures before it increases rapidly and saturates eventually at low temperatures. The inset shows the crystal structure of ZrTe$_5$. The triangle prisms connecting Zr ions and Te1 ions (depicted as the red dashed lines) form 1D chains running along a-axis. Te2 ions, also forming zigzag chains along a-axis, extend along c-axis to connect ZrTe3 chains, making 2D layers. These 2D layers stack along b-axis via Van der Waals interactions to form the 3D crystal. Panel B: Torque magnetometry data as a function of magnetic field for selected samples at 2.5 K. The inset shows $\tau/$H (torque divided by magnetic field). The torque $\tau$ shows $\sim $ quadratic behavior, implying the dominant contribution is either paramagnetic or diamagnetic response. $\tau/$H has no anomaly at low fields except for sample ZQ4, implying no magnetic order. For sample ZQ4, although small anomaly is found for $\tau/$H below $\sim 0.3$ T, the onset of the AHE is $\gtrsim 1$~T, so their origins are different. Furthermore, while every sample shows AHE, not all samples show the anomaly, again implying the different origins between them. Panel C shows ARPES results at T = 17 K. ARPES data show the carrier type is hole-like. The photon energy is 6 eV.
}
\end{figure*}

ZrTe$_5$ has an orthorhombic layered structure with space group Cmcm (D$^{17}_{2h}$)~\cite{XiDaiZrTe5}, as shown in the inset of Fig.~\ref{RT}A. The ZrTe6 triangular prisms (depicted as the red dashed lines) form 1D chains of ZrTe$_3$ running along the $a$-axis. These ZrTe$_3$ chains are connected by additional Te ions which also form zigzag chains along the $a$-axis and extend along the $c$-axis. As a result, they form 2D layers and such 2D layers stack along the $b$-axis via Van der Waals interactions to form the 3D crystal. The Van der Waals interaction along the $b$-axis is very small with its value comparable to graphite~\cite{XiDaiZrTe5}. Therefore, both the 2D single layer and the 3D bulk crystals of ZrTe$_5$ are interesting. A monolayer of ZrTe$_5$ is theoretically predicted to be the quantum spin Hall (QSH) state and 3D bulk ZrTe$_5$ is predicted to lie near the boundary of a weak topological insulator (WTI) and a strong topological insulator (STI)~\cite{XiDaiZrTe5}. In our experiments, 3D bulk crystals are studied.

The transport properties of ZrTe$_5$ were investigated with the current applied along the chain axis ($a$-axis).
Fig.~\ref{RT}A shows the curves of resistivity versus temperature for selected samples showing that the resistivity increases down to the lowest temperatures, where it starts to saturate. The resistivity versus temperature curves in the literature~\cite{TiZrTe5,LnZrTe5} show resistivity profiles with maxima sitting at $T = T_0$, where the system changes its carrier type from hole-like to electron-like~\cite{ZhangZrTe5}. The value of $T_0$ varies in different studies, and can take say $T_0 = 135$~K~\cite{PressureZrTe5} or $65$~K~\cite{LiZrTe5}. It has been shown that the values of $T_0$ can be systematically decreased to lower temperatures via chemical pressure induced by substitution of rare earth elements~\cite{LnZrTe5}. Our samples can be regarded as the situation of $T_0 \lesssim 5$~K. This is consistent with the ARPES data for our samples shown in panel C of Fig.~\ref{RT}, which show that the system is hole-like even at T = 17 K. Although the resistivity increases as the temperature decreases, the value at 5~K is only $\lesssim 5$~m$\Omega$~cm, so the system is still in the metallic regime. 

Since our interest is in the AHE coming from the Berry curvature generated from Weyl nodes, it is worth checking that the AHE is not induced by a conventional mechanism such as ferromagnetism. For this purpose, we have performed torque magnetometry measurements whose results are shown in Fig.~\ref{RT}B for selected samples. No magnetic ordering is observed, confirming that the AHE in ZrTe$_5$, shown in Figs.~\ref{Z2},~\ref{3DBerry},~\ref{ZQ4} (see below), does not come from an ordinary mechanism such as ferromagnetism. This is a natural observation, considering that ZrTe$_5$ consists of no magnetic elements. 

The first clue that implies the existence of Berry curvature in the ZrTe$_5$ system coming from Weyl nodes originates in the observation of the anomalous Hall signals in the $ab$-plane, shown in Fig.~\ref{Z2}. As shown in Fig.~\ref{Z2}A, the negative LMR is narrowly confined to $\lesssim 1\degree$, reproducing the results obtained in Ref.~\cite{LiZrTe5}. Interestingly, the Hall signals show sharp zigzag shaped profile, as shown in Fig.~\ref{Z2}B, suggestive of the existence of the AHE emerging from the Weyl nodes. In order to extract the contribution from the Berry curvature, we have subtracted the linear background at high fields and plotted the anomalous part in panel D. Here we note that while the ARPES data show the carrier type is hole-like, the sign of $\Delta\rho_{yx}$ is ``electron-like'', inferring the contribution of Berry curvature to the AHE. The angular dependence of the magnitude of the anomalous Hall contribution as plotted in panel C shows that the Berry curvature develops rapidly as soon as the magnetic field is tilted away from the $a$-axis, indicative of the sharp sensitivity of the Weyl nodes to the direction of applied magnetic fields. It is clear from Fig.~\ref{Z2} that the amplitude of the anomalous contribution is not symmetric with respect to the angle for sample Z2. This suggests that the observed AHE in this case has an additional contribution which emerges from the in-plane component of the magnetic field that was not cancelled completely due to a slight sample misalignment. 

In order to extract the contribution to the AHE from the in-plane magnetic field as well as to separate its contribution from the out-of-plane component, a double axis rotator was employed to obtain the in-situ 3D full $4\pi$ solid anglular dependence of the anomalous Hall signal, which is plotted as a function of two angles, $\varphi$ and $\theta$, respectively representing the azimuth and elevation angle, as shown in Fig.~\ref{3DBerry} for sample ZQ3. Panels j, k, and l of Fig.~\ref{3DBerry} show a spherical plot of the anomalous Hall component, with the radius representing the magnitude of the anomalous Hall signal, or, effectively the strength of the Berry curvature. When carefully aligned using the double axis rotator, the out-of plane ($ab$-plane) anomalous Hall signals show completely antisymmetric behavior with respect to the $a$-axis, as shown in panels a, b, and c.

Next, we pay attention to the contribution coming from the in-plane (ac-plane) magnetic field.
Remarkably, large, sharp in-plane anomalous Hall signals are observed for every angle except when the magnetic field is parallel to the $a$-axis, as shown in panels d, e, and f. The anomalous Hall contribution now has less sensitivity to the angle compared to the case of the $ab$-plane, with the anomalous Hall amplitude saturating only gradually as the angle increases. The in-plane Hall effect is quite anomalous, as the conventional Lorentz force cannot give to any Hall signal in this orientation. Yet, the experimental results show that there is a large contribution of the AHE, comparable or even larger than that coming from out-of plane ($ab$-plane), strongly suggesting the contribution from Berry curvature produced by Weyl nodes.

When the magnetic field is rotated in the $bc$-plane, however, the contribution from the out-of plane ($ab$-plane) and in-plane ($ac$-plane) mix with each other and the AHE shows mixed behavior - as shown in panels g, h, and i. Interestingly, if the anomalous Hall signals are antisymmetrized with respect to the angle (curve in dark cyan in panel i), they give similar results as obtained for the case of out-of plane rotation ($ab$-plane). This suggests that the contribution from the $b$-axis can be simply obtained as the projection of the magnetic field onto the b-axis. By contrast, the in-plane ($ac$-plane) contribution cannot be obtained by the simple symmetrization process (curve in olive in panel i). In other words, a simple projection of magnetic field onto the c-axis cannot reproduce the results. This infers that there might be two kinds of Weyl pairs which respond differently to external magnetic fields.  

In order to further test that the AHE is coming from the Weyl nodes, the temperature dependence of the AHE was investigated, with the results shown in Fig.~\ref{ZQ4}. As is clear from panel A, the negative LMR, suggestive of the chiral anomaly coming from the Weyl nodes, starts to become prominent below $\sim 60$~K. The anomalous Hall signal shows similar trend, which is clear from panel B, becoming substantial below $\sim 60$~K. This implies the close relationship between the negative LMR and the AHE, confirming further that the origin of AHE is coming from the Weyl nodes.  

We briefly discuss the results of the anomalous Nernst effect (ANE) shown in panels C and D of Fig.~\ref{ZQ4}. Since the AHE is observed for ZrTe$_5$, it is natural to expect that the system also shows an ANE. One advantage of the Nernst effect is that its sensitivity is higher than the Hall signal~\cite{Behnia2007,Behnia2011,Behnia2015}. Panels C and D of Fig.~\ref{ZQ4} show the angular dependence of the AHE and ANE for sample Z5. The close relation implies the same origin for the AHE and the ANE.

In conclusion, a clear AHE develops for every sample of ZrTe$_5$ under both in-plane ($ac$-plane) and out-of plane ($ab$-plane) magnetic fields. This AHE emerges at the same temperature where the negative LMR evolves, implying their common origin. Combined with the observation of torque magnetometry showing that no magnetic ordering develops in ZrTe$_5$, this suggests the existence of Weyl nodes in the system. The observation of a sharp ANE further strengthens this claim. However, there are no theoretical predictions showing that Weyl nodes should exist in ZrTe$_5$. Furthermore, ARPES measurements suggest that a small gap is open at $\Gamma$ point for ZrTe$_5$ samples with $T_0 = 135$~K~\cite{ ManzoniZrTe5, ShenZrTe5}. One potential scenario for Weyl nodes appearing in ZrTe$_5$ is that since the energy gap at the $\Gamma$ point is small, Weyl nodes can be induced when TRS is broken by large Zeeman energy under an applied magnetic field, which splits the bands at $\Gamma$ point. The other possibility is that since ZrTe$_5$ is located near the boundary of a WTI and a STI~\cite{XiDaiZrTe5}, if the inversion symmetry is broken in the system, then with a little bit of change in the lattice parameter, the system automatically falls into a Weyl semimetallic phase, according to the general phase diagram proposed by Murakami \etal~\cite{Murakami2007,Murakami2008,Murakami2011,Murakami2014}. The fact that the anomaly of the resistivity happens at $T_0\lesssim 5$ K for our ZrTe$_5$ samples is suggestive of a subtle difference in doping or lattice constant in our samples compared to samples with $T_0 = 135$~K. Because the inversion symmetry broken Weyl states contains at least four Weyl nodes, or two kinds of Weyl pairs, it would be interesting to investigate how they respond to an applied magnetic field. Our results indicate that investigating whether the inversion symmetry is broken as well as how the Weyl nodes appear in ZrTe$_5$ are interesting directions to pursue in future research.

\newpage

%%%%%%%%%%%%%%%%%%%%%%%%%%%%%%%%%%
%%%%%%%%%%%%%%%%%%%%%%%%%%%%%%%%%%
%%%%%%%%%%%%%%%%%%%%%%%%%%%%%%%%%% FIGURE 1

%%%%%%%%%%%%%%%%%%%%%%%%%%%%%%%%%
%%%%%%%%%%%%%%%%%%%%%%%%%%%%%%%%%
%%%%%%%%%%%%%%%%%%%%%%%%%%%%%%%%% FIGURE 2
\begin{figure*}[t]
\includegraphics[width=15 cm]{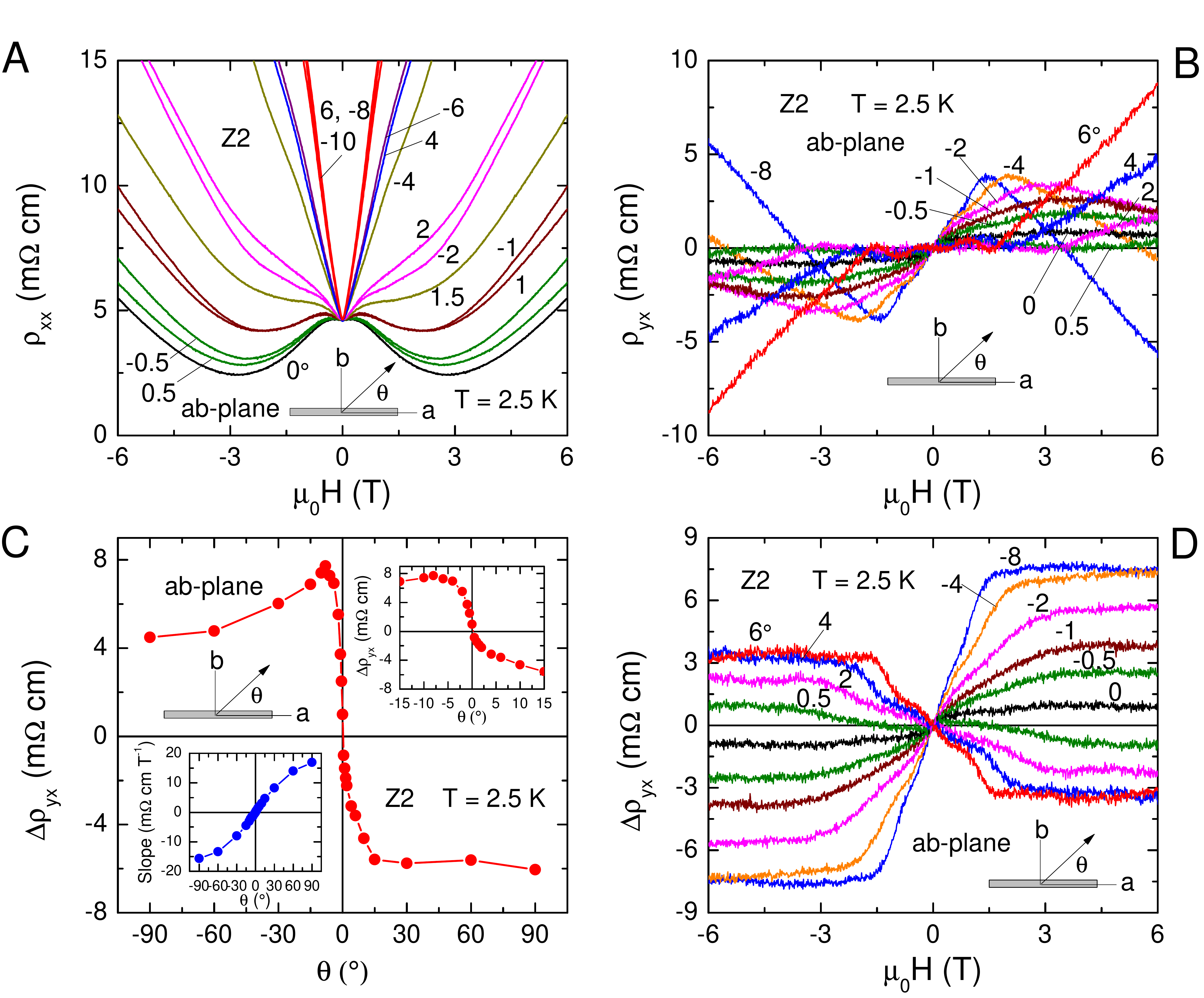}
\caption{\label{Z2} 
Angular dependence of MR and Hall at selected angles in ab-plane for sample Z2. Panel A: Negative LMR is observed in small angle regime $\theta \lesssim$ 1 $^\circ$. Panel B: Hall signals show zigzag shape manifesting the existence of anomalous Hall contribution. Panel C: Full angular dependence of the anomalous Hall amplitude obtained from Panel D. The right top inset shows the anomalous Hall amplitude at small angles and the left lower inset shows the angular dependence of the background slope. Panel D: Anomalous Hall signals after linear background subtraction from Panel B.  
}
\end{figure*}

%%%%%%%%%%%%%%%%%%%%%%%%%%%%%%%%%%
%%%%%%%%%%%%%%%%%%%%%%%%%%%%%%%%%%
%%%%%%%%%%%%%%%%%%%%%%%%%%%%%%%%%% FIGURE 3
\begin{figure*}[t]
\includegraphics[width=15 cm]{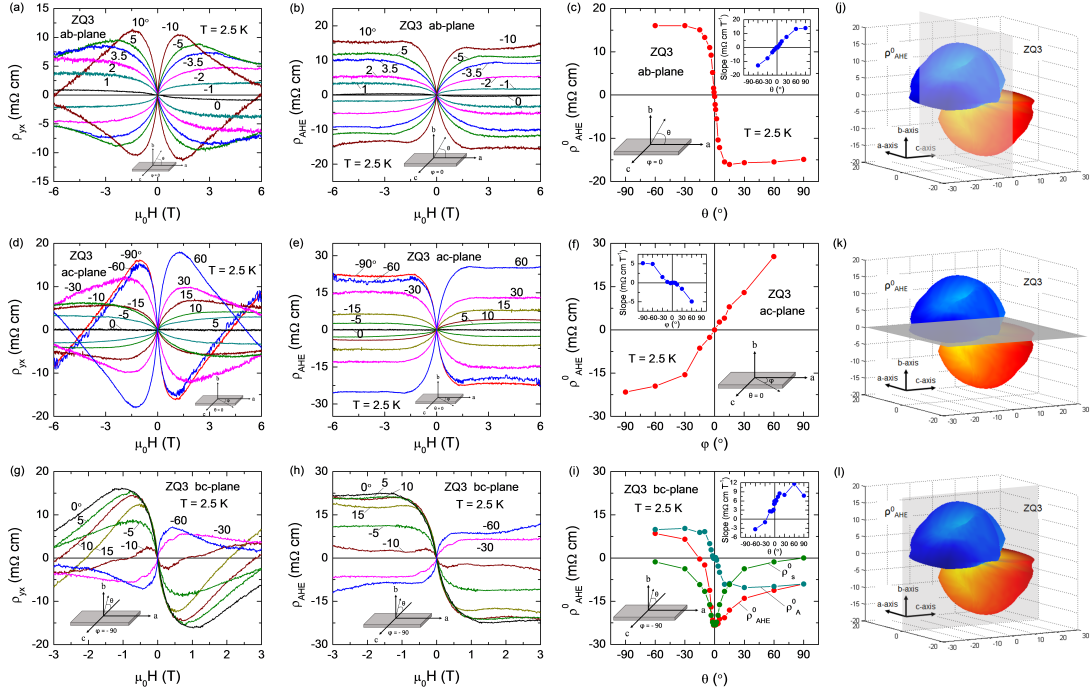}
\caption{Full 3D angular dependence of the AHE $\rho^0_{\mathrm{AHE}}$ investigated for sample ZQ3. When the magnetic field is rotated in $ab$-plane (out of plane), sharp anomalous Hall signals are observed with the Berry curvature saturating rapidly as the magnetic field tilts away from $a$-axis (panels a, b, c). When the magnetic field is rotated instead in $ac$-plane (in-plane), comparable or even larger anomalous Hall signals are observed, now with the Berry curvature less sensitive to the direction of the magnetic field (panels d, e, f). When the magnetic field is rotated in $bc$-plane (panels g, h, i), the system picks up the mixed contribution of anomalous Hall signals showing tilted behavior of the Berry curvature as a function of angle as shown in panel i. The antisymmtrized contribution $\rho_A^0$ shown as the curve in dark cyan resembles the Berry curvature obtained for $ab$-plane. By contrast, the symmetrized contribution $\rho_S^0$ does not follow the profile shown in panel f for the case of $ac$-plane, inferring that there might exist two different types of Weyl pairs in the system. Panels j, k, l show the 3D spherical plot of the anomalous Hall signal $\rho^0_{\mathrm{AHE}}$, with the blue color representing the negative value and orange showing the positive value. The principal planes are also shown for clarity. \label{3DBerry} 
}
\end{figure*}

%%%%%%%%%%%%%%%%%%%%%%%%%%%%%%%%%%
%%%%%%%%%%%%%%%%%%%%%%%%%%%%%%%%%%
%%%%%%%%%%%%%%%%%%%%%%%%%%%%%%%%%% FIGURE 4
\begin{figure*}[t]
\includegraphics[width=15 cm]{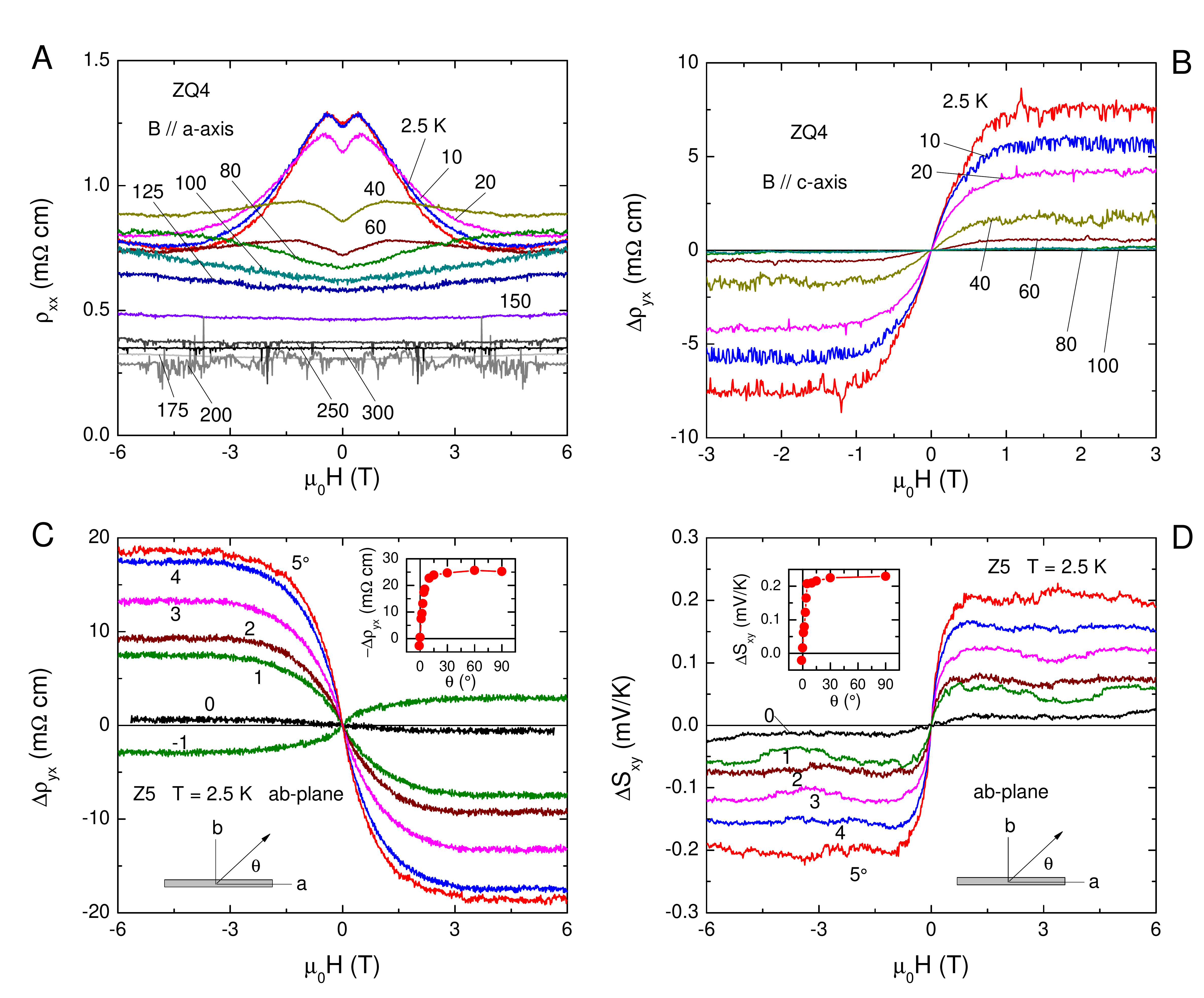}
\caption{\label{ZQ4}
Temperature dependence of MR and Hall signals for sample ZQ4 (panels A, B). Panel A: The negative LMR starts to grow rapidly below $\sim $ 60 K. Panel B: Correspondingly, the anomalous Hall signals become prominent below $\sim $ 60 K, showing close relation between the negative LMR and the AHE.
Angular dependence of anomalous Hall (panel C) and Nernst (panel D) signals for sample Z5. The insets show the angular dependence of the amplitude of the anomalous signals. The close similarity between the anomalous Hall and Nernst effect suggests the same origin of both signals.
}
\end{figure*}

%\bibliographystyle{naturemag_noURL}
%\bibliography{ZrTe5_Ref.bib}

\end{document}